\documentclass[prl,showpacs,twocolumn]{revtex4}
\usepackage{graphicx} 
\usepackage{dcolumn}  
\usepackage{bm}       
\usepackage{amsmath}
\usepackage{amsfonts}
\usepackage{amssymb}

\begin{document}

\title{Turbulent luminance in impassioned van Gogh paintings}

\author{J.L. Arag\'on}
\affiliation{Centro de F\'{\i}sica Aplicada y Tecnolog\'{\i}a
  Avanzada, Universidad Nacional Aut\'onoma de M\'exico, Apartado
  Postal 1-1010, Quer\'etaro 76000, M\'exico.}

\author{Gerardo G. Naumis}
\affiliation{Instituto de F\'{\i}sica, Universidad Nacional Aut\'onoma
  de M\'exico, Apartado Postal 20-364, 01000 M\'exico, Distrito
  Federal.}

\author{M. Bai}
\affiliation{Laboratorio de F\'{\i}sica de Sistemas Peque\~nos y
  Nanotecnolog\'{\i}a, Consejo Superior de Investigaciones
  Cient\'{\i}ficas, Serrano 144, 28006 Madrid, Spain.}

\author{M. Torres}
\affiliation{Instituto de F\'{\i}sica Aplicada, Consejo Superior de
  Investigaciones Cient\'{\i}ficas, Serrano 144, 28006 Madrid, Spain.}

\author{P.K. Maini}
\affiliation{Centre for Mathematical Biology, Mathematical Institute,
  24-29 St Giles Oxford OX1 3LB, U.K.}


\maketitle

\textbf{Everything in the last period of Vincent van Gogh paintings
  seems to be moving; this dynamical style served to transmit his own
  feelings about a figure or a landscape. Since the early
  impressionism, artists emprically discovered that an adequate use of
  luminance could generate the sensation of motion. This sentation was
  more complex in the case of van Gogh paintings of the last period:
  turbulence is the main adjective used to describe these
  paintings. It has been specifically mentioned, for instance, that
  the famous painting \emph{Starry Night}, vividly transmits the sense
  of turbulence and was compared with a picture of a distant star from
  the NASA/ESA Hubble Space Telescope, where eddies probably caused by
  dust and gas turbulence are clearly seen \cite{Daily}. It is the
  purpose of this paper to show that the probability distribution
  function (PDF) of luminance fluctuations in some impassioned van
  Gogh paintings, painted at times close to periods of prolonged
  psychotic agitation of this artist, compares notable well with the
  PDF of the velocity differences in a turbulent flow as predicted by
  the statistical theory of Kolmogorov. This is not the first time
  that this analogy with hydrodinamic turbulence is reported in a
  field far different from fluid mechanics; it has been also observed
  in fluctuations of the foreign exchange markets time series
  \cite{Ghashghaie}.}

Luminance is a measure of the luminous intensity per unit area. It
describes the amount of light that passes through or is emitted from a
particular area, and falls within a given solid angle
\cite{Luminance}. Its psychological effect is bright and thus is an
indicator of how bright a surface will appear. Luminance has been used
by artists to produce certain effects. For instance, the technique of
equiluminance has been used since the early impressionism to transmit
the sensation of motion in a painting. Notably Claude Monet in his
famous painting \emph{Impression, Sunrise} used regions with the same
luminance, but constrasting colors, to make his sunset twinkle. The
biological basis behind this effect is that color and luminance are
analyzed by different parts of the visual system; shape is registerd
by the region that process color information but motion is registered
by the colorblind part. Thus, albeit equiluminant regions can be
differentiated by color contrast, they have poorly defined positions
and it may seem to vibrate \cite{Livingstone}. It seems likely that
van Gogh dominated this technique but some of the paintings of his
last period produce a more disturbing feeling: they transmit the sense
of turbulence. By assuming that luminance is the property that van
Gogh used to transmit this feeling (without being aware of
it), we will quantify the turbulence of some impassioned paintings by
means of a statistical analysis of luminance, similar to the
statistical approach that Andrei Kolmogorov used to study fluid
turbulence.

We mainly study van Gogh's \emph{Starry Night} (June 1889), which
undoubtedly transmits the feeling of turbulence. Also, as samples of
another turbulent pictures, we analyze \emph{Wheat Field with Crows}
(July, 1890, just before van Gogh shot himself),  \emph{Road with
  Cypress and Star} (May, 1890) and \emph{Self-portrait with pipe and
  bandaged ear} (January, 1889). By considering the analogy with the
Kolmogorov turbulence theory, from our results we can conclude that
the turbulence in luminance of the studied van Gogh paintings is like
real turbulence. Thus, the distribution of luminance on thes paintings
exhibit the same characteristif features of a turbulent fluid. This
resul may offer a clue on why van Gogh was capable of transmitting the
feeling of turbulence with high realism. Our results also reinforce
the idea that scientific objectivity may help to determine the
fundamental content of artistic paintings, as was already done with
Jackson Pollock's fractal paintings \cite{Taylor,Mureika}.  Along this
same ideas, it also worthy to mention that another notable ability of
van Gogh was recently remarked with an experiment with bumblebees that
had never seen natural flowers; insects were more attracted by van
Gogh's \emph{Sunflowers} than by other paintings containing flowers
\cite{Chittka}. From this observation, Chittka and Walker suggest that
van Gogh's flower paintings have captured the essence of floral
features from a bee's point of view.

\begin{figure}[!t]
\begin{center}
\includegraphics[width=8.0cm]{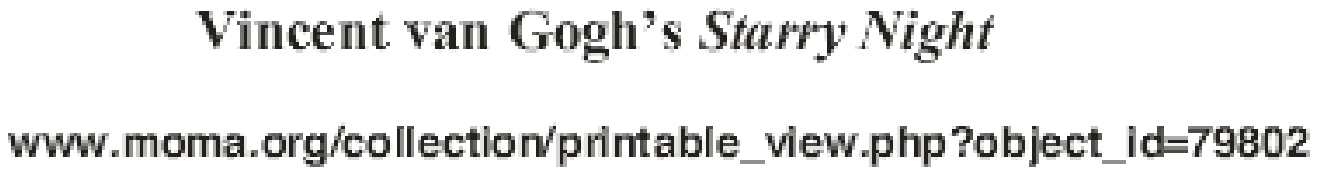}
  \caption{Vincent van Gogh's \emph{Starry Night}.}
\label{fig:fig1}
\end{center}
\end{figure}

The statistical model of Kolmogorov \cite{Kolmogorov1, Kolmogorov2} is
a foundation for modern turbulence theory. The main idea is that at
very large Reynolds numbers, between the large scale of energy input
($L$) and the dissipation scale ($\eta$), at which viscous frictions
become dominant, there is a myriad of small scales where turbulence
displays universal properties independent of initial and boundary
conditions. In particular, in the inertial range Kolmogorov predicts a
famous scaling property of the second order structure function, $S_2 (
{\bf R} ) = \langle \left( \delta v _R \right)^2 \rangle$, where
$\delta v _R = v ({\bf r} + {\bf R}) - v ({\bf r})$ is the velocity
increment between two points separated by a distance ${\bf R}$ and $v$
is the component of the velocity in the direction of ${\bf R}$. In his
first 1941 paper \cite{Kolmogorov1} Kolmogorov postulates two
hypotheses of similarity that led to the prediction that $S_2 ( {\bf
  R})$ scales as $\left( \varepsilon R \right) ^{2/3}$, where $R = \|
{\bf R} \|$ and $\varepsilon$ is the mean energy dissipation rate per
unit mass.  Under the same assumptions, in his second 1941 turbulence
paper \cite{Kolmogorov2} Kolmogorov found an exact expression for the
third moment, $\langle \left( \delta v_R \right)^3 \rangle$, which is
given by $S_3 ( {\bf R}) = -\frac{4}{5} \varepsilon R$. And even more,
he hypothesized that this scaling results generalizes to structure
functions of any order, \emph{i.e.} $ S_n ({\bf R}) = \langle \left(
  \delta v_R \right)^n \rangle \propto R ^{\xi_n}$, where $\xi_n =
n/3$.  Experimental measurements show that Kolmogorov was remarkably
close to the truth in the sense that statistical quantities depend on
the length scale $R$ as a power law. The intermittent nature of
turbulence causes, however, that the numerical values of $\xi_n$
deviate progressively from $n/3$ when $n$ increases, following a
concave curve below the $n/3$ line \cite{Warhaft}. In 1962, Kolmogorov
\cite{Kolmogorov3} and Obukhov \cite{Obukhov} recognize that
turbulence is too intermitent to be described by simple power laws and
propose a refinement that yields a log-normal form of the probability
density of $\varepsilon$.

An important function to characterize turbulence is the PDF of
velocity differences $\delta v_R$, and different models have been
proposed to describe the shape of this function at different scales
$R$. We adopt here the approach by Castaign \emph{et al.}
\cite{Castaign} that, supported by experimental results, follows the
idea of the log-normal form of $\varepsilon$. By superimposing several
Gaussians at different scales, it is inferred that the shape of the
PDF goes from nearly Gaussian at large scales $R$ to nearly
exponential at small scales. The number of superimposed Gaussians is
controlled by a parameter, $\lambda$, which is the only parameter that
must be fitted to the data.  A large value of $\lambda$ means that
many scales contribute to the results, and thus the PDF develops tails
that decay much slower than a pure Gaussian correlation.

\begin{figure}[!t]
\begin{center}
\includegraphics[width=8.0cm]{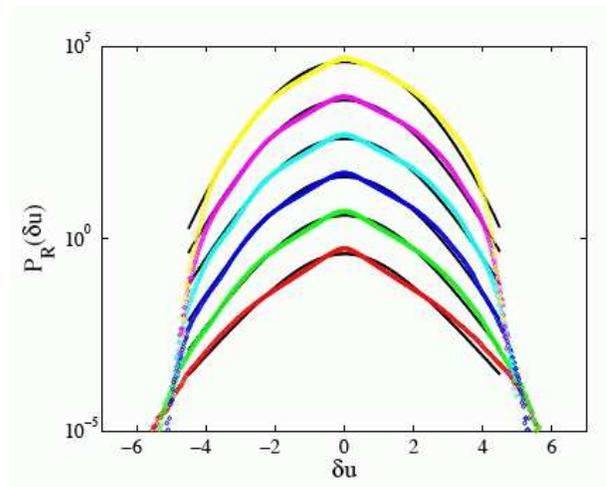}
  \caption{Semilog plot of the probability density $P_R (\delta u)$ of
    luminance changes $\delta u$ for pixel separations $R=60$, $240$,
    $400$, $600$, $800$, $1200$ (from bottom to top). Curves have been
    vertically shifted for better visibility. Data points were fitted,
    according to Ref. \cite{Castaign}, and the results are shown in
    full lines; parameter values are $\lambda = 0.2$, $0.15$, $0.12$,
    $0.11$, $0.09$, $0.0009$ (from bottom to top).}
\label{fig:fig2}
\end{center}
\end{figure}

\emph{Starry Night}, painted during his one year period in the Saint
Paul de Mausole Asylum at Saint-R\'{e}my-de-Provence, is undoubtedly
one of the best known and most reproduced paintings by van Gogh
(Fig. \ref{fig:fig1}). The composition describes an imaginary sky in a
twilight state, transfigured by a vigorous circular brushwork.  To
perform the luminance statistics of \emph{Starry Night}, we start from
a digitized, $300$dpi, $2750 \times 3542$ image obtained from The
Museum of Modern Art in New York (where the original paint lies),
provided by Art Resource, Inc.  In a digital image, the luminance of a
pixel is obtained from its RBG (red, green and blue) components as
\cite{Gonzalez} $0.299 R + 0.587 G + 0.114 B$. This approximate
formulae takes into account the fact that the human eye is more
sensitive to green, then red and lastly blue. Thus, we calcuate the
PDF of pixel luminance fluctuations by building up a matrix whose rows
contain difference in luminance $\delta u$ and columns contain
separation between pixels $R$. From this matrix, we determine the
normalized PDF of the luminance differences $P_R (\delta u _R) =
\delta u _R / ( \langle ( \delta u _R ) ^2 \rangle ) ^{1/2}$. In
Figure \ref{fig:fig2} we show this function for six pixel separations,
$R=60$, $240$, $400$, $600$, $800$, $1200$.  In order to rule out
scaling artifacts, we have systematically recalculate the PDF function
to images with lower resolutions (with an adequate rescaling of the
pixel separations $R$). No significative differences appear up to
images with resolutions lower that $150 \times 117$ pixels, where the
details of the brushwork are lost.

\begin{figure}[t!]
\begin{center}
\includegraphics[width=8.0cm]{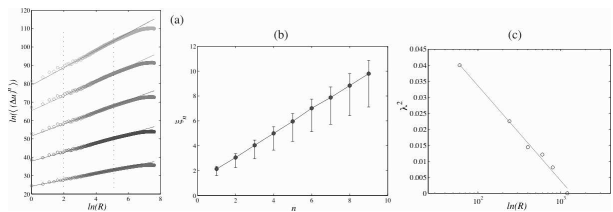}
  \caption{(a) Log-log plot of the statistical moments $\langle \left(
      \delta u (R) \right) ^n \rangle$, with $n = 1$, $2$, $3$, $4$,
    $5$ (from bottom to top). The notation $ln$ denotes a natural
    logarithm and in each case the stright line indicates the
    least-squares fit to the range of scales limited by the two dashed
    lines in the plot; (b) exponent $\xi_n$ of the statistical moments
    as a function of $n$. For a given $n$, the exponent and error bar
    was caclulated by the method proposed in Ref. \cite{Mitra}; (c)
    Dependence of $\lambda ^2$ on $R$. Data points are fitted to a
    straight line by a least-square method.}
\label{fig:fig3}
\end{center}
\end{figure}

We can take the analogy with fluid turbulence further. By considering
the large length scale as $L=2000$ pixels, which is size of the
largest eddy observed in the \emph{Starry Night}, in Fig. 3a, we show
a log-log plot of the statistical moments with $n = 1$, $2$, $3$, $4$,
$5$ (from bottom to top), that show power-law regimes. In each case
stright line indicates the least-squares fit to the range of scales
limited by the two dashed lines in the plot.  In Fig. \ref{fig:fig3}b,
the scaling exponent $\xi_n$, of the first ten statistical moments are
shown as a function of $n$. Albeit data point can be fitted with great
accuracy to a straight line (implying a simple scaling consistent with
a self-similar picture of turbulence but no with intermittence),
scaling exponents show deviations from the self-simmilar values as
indicated by the error bars. To determine scaling exponents and error
bars, we follow the method proposed in Ref. \cite{Mitra}, based on local
slopes.

The PDF of luminance, for a given $R$, shown in Fig \ref{fig:fig2}
were fitted according to the model by Castaign \emph{et al.}
\cite{Castaign} yielding a notably good fit. Results are shown in
the same figure with full lines; parameter values are $\lambda = 0.2$,
$0.15$, $0.12$, $0.11$, $0.09$, $0.0009$ (from bottom to top).

Finally, Figure \ref{fig:fig3}c shows the dependence of $\lambda ^2$ on
$\ln (R)$ for the first six moments. As expected, the parameter
$\lambda ^2$, which measures the variance of the log-normal
distribution \cite{Castaign}, decreases linearly with $\ln (R)$

\begin{figure}[!t]
\begin{center}
\includegraphics[width=7.0cm]{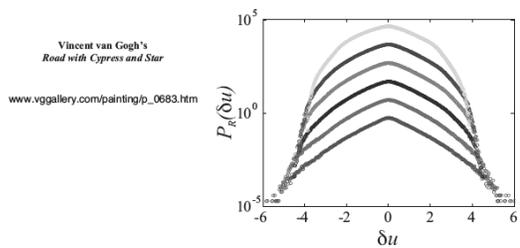}
\caption{\emph{Wheat Field with Crows} (top) and its PDF (bottom) for
  pixel separations $R=10$, $40$, $80$, $150$, $250$, $350$ (from bottom to
  top).  The studied image was taken from the Van Gogh Museum, Amsterdam,
  webpage.}
\label{fig:fig5}
\end{center}
\end{figure}

\begin{figure}[!t]
\begin{center}
\includegraphics[width=8.0cm]{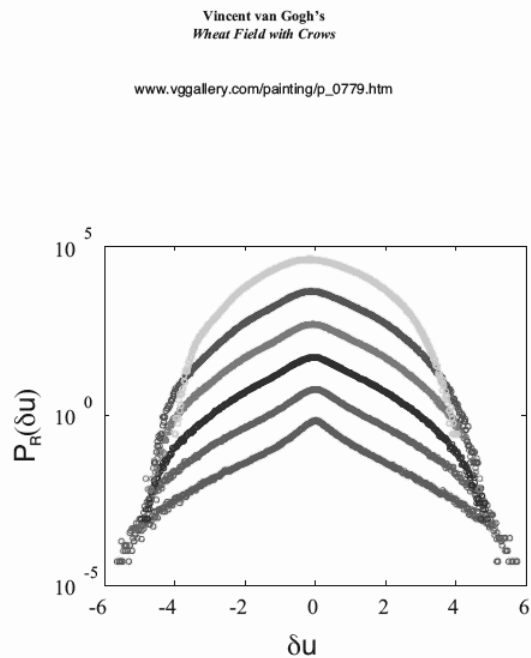}
  \caption{Left: \emph{Road with Cypress and Star} (Rijksmuseum
    Kr\"oller-M\"uller, Otterlo). Right: PDF for pixel separations
    $R=2$, $5$, $15$, $20$, $30$, $60$ (from bottom to top). The studied image
    was taken from the WebMuseum-Paris, webpage.}
\label{fig:fig6}
\end{center}
\end{figure}

From van Gogh's 1890 period, we firstly analyze \emph{Wheat Field with
  Crows}, which is one of his last paintigs. To perform the luminancwe
statistics we use a digitized, $300$dpi, $5369 \times 2676$ image
obtained from The Van Gogh Museum, in Amsterdam, provided by Art
Resource, Inc. Figure \ref{fig:fig5} shows the PDF for six pixel separations. 
Secondly, we analyzed \emph{Road with Cypress and Star}, that was
painted just after the last and most prolonged psychotic episode of
van Gogh's life, lasting from February to April 1890, during which the
artist suffered terrifying hallucinations and severe agitation
\cite{Blumer}. We use a digitized, $300$dpi, $815 \times 1063$ image
obtained from the WebMuseum, Paris, webpage. Figure \ref{fig:fig6}
shows the PDF for six pixel separations. In both cases the curves show close
simmilarity to the behaiviour of the PDF of fluid turbulence.

\begin{figure}[!t]
\begin{center}
\includegraphics[width=8.0cm]{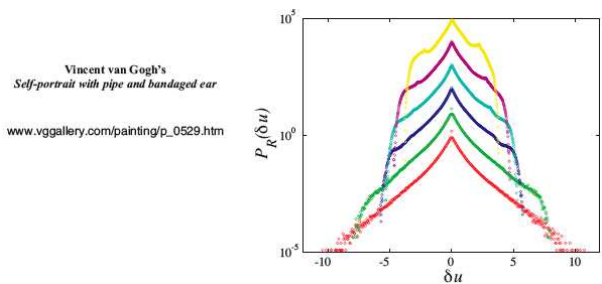}
  \caption{\emph{Self-portrait with pipe and bandaged ear} and its
    PDF for six pixel separations, $R=2$, $5$, $15$, $20$, $30$, $60$ 
    (from bottom to top). The studied image was taken from The Vincent van Gogh 
    Gallery webpage.}
\label{fig:fig7}
\end{center}
\end{figure}

For comparison purposes, in Fig. \ref{fig:fig7} we show van Gogh's
\emph{Self-portrait with pipe and bandaged ear} and its PDF for
six pixel separations. In a well known episode of his life, on 23 December 1888,
Vincent van Gogh mutilated the lower portion of his left ear. He was
hospitalized at the H\^otel-Dieu hospital in Arles and prescribed
potassium bromide \cite{Blumer}. After some weeks, van Gogh recovered
from the psychotic state and, in a stage of absolute calm (as himself
described in a letter to his brother Theo and sister Wilhemina
\cite{Letters}), he painted the self-portrait with pipe. As it can be
seen in Fig. \ref{fig:fig7}, the PDF of this paint departs from what
is expected in Kolmogorov's model of turbulence. We analyzed a
$300$dpi, $605 \times 732$ image obtained from The Vincent van Gogh
Gallery webpage.

In summary, our results show that \emph{Starry Night}, and other
impassioned van Gogh paintings, painted during periods of prolonged
psychotic agitation trasnmited the escence of turbulence with high
realism. We use Kolmogorov's model of turbulence to determine the
degree of "realism" contained in the paintings.  We are also
suggesting new tools and approaches that open the possibility of
quantitative objective research for art representation.

\begin{acknowledgments}
  This work has been partially supported by by DGAPA-UNAM (Grant
  No. IN-108502-3), CONACyT (Grant No. D40615-F) and MCYT-Spain (Grant
  No. FIS2004-03237).
\end{acknowledgments}

\end{document}